\documentstyle[epsf,emulateapj,apjfonts]{article}

\def\simlt{\mathrel{\hbox to 0pt{\lower 3.5pt\hbox{$\mathchar"218$}\hss}
      \raise 1.5pt\hbox{$\mathchar"13C$}}}
\def\simgt{\mathrel{\hbox to 0pt{\lower 3.5pt\hbox{$\mathchar"218$}\hss}
      \raise 1.5pt\hbox{$\mathchar"13E$}}}

\slugcomment{Accepted for publication in the {\sl Astrophysical Journal Letters}}

\begin{document}

\title{How young are early-type cluster galaxies~?~ Quantifying the 
young stellar component in a rich cluster at $z=0.41$.\altaffilmark{1}}
\author{Ignacio Ferreras, Joseph Silk}
\affil{Nuclear \& Astrophysics Lab. 1 Keble Road, Oxford OX1 3RH, United Kingdom}
\authoremail{
1. ferreras@astro.ox.ac.uk \\
2. silk@astro.ox.ac.uk}
\altaffiltext{1}{Work based on observations made with the NASA/ESA 
Hubble Space Telescope, obtained from the data archive at the Space 
Telescope Science Institute. STScI is operated by the Association of 
Universities for Research in Astronomy, Inc. under NASA contract 
NAS 5-26555.}

\vskip0.5truecm

\begin{abstract}
We present a new method of quantifying the mass fraction of young stars
in galaxies by analyzing near-ultraviolet (NUV)$-$optical colors.
We focus our attention on early-type cluster galaxies, whose star 
formation history is at present undetermined. Rest frame NUV (F300W) and 
optical (F702W) images of cluster Abell 851 ($z=0.41$) using HST/WFPC2 
allow us to determine a NUV$-$optical color-magnitude relation, whose 
slope is incompatible with a monolithic scenario for 
star formation at high redshift. A degeneracy between a young stellar 
component and its fractional mass contribution to the galaxy is found, 
and a photometric analysis comparing the data with the predictions for
a simple two-stage star formation history is presented. The analysis
shows that some of the early-type galaxies may have fractions 
higher than 10\% of the total mass content in stars formed at $z\sim 0.5$. 
An increased scatter is found in the color-magnitude relation at
the faint end, resulting in a significant fraction of faint blue
early-type systems. This would imply that less massive galaxies undergo
more recent episodes of star formation, and this can be explained
in terms of a positive correlation between star formation efficiency 
and luminosity.
\end{abstract}



\section{Introduction}

Dating the stellar component in galaxies is one of the most important
issues in understanding the process of galaxy formation and morphological
segregation. Early-type systems can be explained dynamically by coalescence 
of galaxies of similar masses. In the standard hierarchical merging
scenario these events should take place at late stages. However, 
the first photometric analyses of this type of galaxy hinted at 
old stellar populations. These seemingly contradictory conclusions
can be reconciled as long as the star formation and the dynamical 
histories are decoupled, so that stars form first and the latest
mergers involve only stars and hot gas. An accurate determination
of the star formation history is hindered by degeneracies 
which are sources of controversy with respect to estimates of ages 
of the stellar populations (e.g. Kuntschner 2000, Trager et al. 2000). 
The quest is still on for a grid of observables which can disentangle 
the effects of age and metallicity. Balmer indices and mass-to-light 
ratios are the best ``stellar clocks'' at present. We propose an 
alternative age estimator, namely the flux around rest 
frame $\lambda\sim 2000$\AA\  which is highly sensitive to recent 
star formation. 
The Near Ultraviolet (NUV) spectral range has been targeted in local
early-type galaxies as an indicator of very old ages through the
population of low-mass core-Helium burning stars (Brown et al. 1997; 
Yi et al. 1999), whose NUV flux peaks around $\lambda\sim
1500$\AA\ . However, the flux from a small fraction of 1~Gyr 
main-sequence stars will overwhelm this contribution.
Furthermore, targets at moderate redshift ($z\sim 0.5$) cannot be 
older than $t\simlt 10$ Gyr for a reasonable cosmology, reducing
the importance of old stellar populations in NUV. 

In this paper we perform a combined NUV and optical analysis of 
early-type systems in cluster Cl0939+4713 ($z=0.41$), comparing the 
photometry with a simple two-stage star formation scenario. 
Cl0939+4713 is the most distant Abell cluster (Abell 851) and 
one of the richest clusters known. It features a large fraction
of post-starburst (E+A or k+a) galaxies (Belloni et al. 1995)
and its X-ray structure hints at the merging of two clusters
(Schindler \& Wambsganss 1996). Hence, this cluster is an ideal 
first target for the analysis of recent episodes of star formation 
in early-type cluster galaxies.

\section{Data Reduction and Photometry}
The images were retrieved from the archive of the Hubble Space 
Telescope. Both were obtained with the Wide Field and Planetary
Camera 2 during cycle 5 (F300W, 14,000 sec) and cycle 4
(F702W, 21,000 sec). References for the projects associated
with these images can be found in Buson et al. (2000, F300W) and
Dressler et al. (1994, F702W), respectively. The NUV images 
correspond to the spectral range around rest frame 2000\AA\  at 
the redshift of the cluster, whereas F702W roughly maps 
rest frame $V$ band. 
The optical (F702W) images of cluster Cl0939+4713 were used by
the MORPHS collaboration (Smail et al. 1997) who estimated the
morphologies of a sample of clusters at moderate redshifts. We
thereby use their classification in order to extract the sample
of early-type systems. All of the galaxies used in our sample
were classified as types E, S0 or E/S0.

The pipeline-reduced images retrieved from the archive of the
Space Telescope Science Institute were separately registered
chip-by-chip and added using a clipping algorithm 
(IRAF/sigclip) which eliminated cosmic rays without altering 
the photometry. 
The final source detection and
photometric output was performed using SEXtractor 
(Bertin \& Arnouts 1996). Instead of using a fixed aperture, 
we decided to use SEXtractor's ``best'' estimate for the magnitude, 
which starts with an 

\centerline{\null}
\vskip3truein
\includegraphics{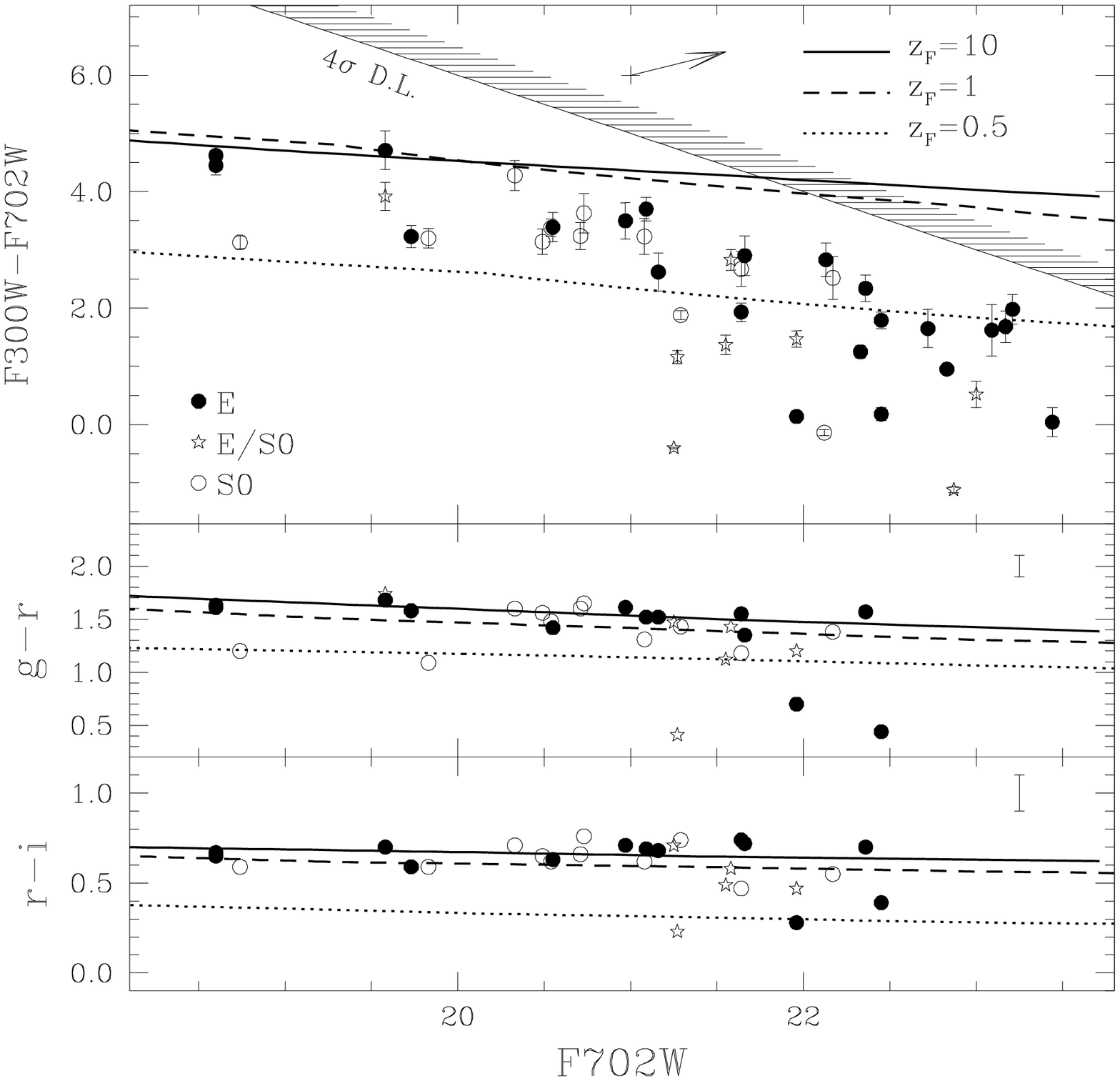}
\figcaption[nuvapj_f1.eps]{Color-magnitude relations of early-type galaxies
in cluster Cl0939+4713 ($z=0.41$). The solid, dashed and dotted lines
correspond to a dustless pure metallicity sequence of galaxies (i.e. all 
of the same age but varying metallicity) formed at redshifts of $z_F=10,1$ 
and $0.5$, respectively, using the observed color-magnitude relation
of local clusters as constraint. The arrow represents the evolution in 
the color and magnitude if we add dust according to the prescription of
Calzetti (1997) for starbursts. A typical error bar for $g-r$ and $r-i$
colors is shown to the right of both diagrams. The shaded line gives
the $4\sigma$ detection limit from the shallower F300W images.\label{f1}}
\vskip+0.2truein

\noindent
adaptive aperture and checks whether nearby
sources are expected to bias the magnitude by more than 0.1 mag,
in which case an isophotal estimate is chosen instead. This method
allows us to get an estimate for the flux that is closest to
the actual ``total'' magnitude. Our F702W magnitudes 
agree with Smail et al. (1997) to better than 0.1 mag but
differ significantly from Buson et al. (2000) who used a fixed 
$1.2\arcsec$ aperture. Additional optical photometry in $g$ and $r$ 
was retrieved from Dressler \& Gunn (1992). Out of the 72 galaxies 
classified as early-type systems in the F702W images, we detected 
42 {\sl both} in F300W and F702W out of which 30 
(down to F702W$\sim 22$) are listed in Dressler \& Gunn (1992).

Figure~1 shows the NUV$-$optical and optical CM relations. The
latter are compatible with a single burst at high redshift as shown 
in the figure by the thick solid and dashed lines, which give the CM
of a simple stellar population formed at redshifts of $z_F=10$
and $1$, respectively. The prediction for a more recent burst at
$z_F=0.5$ is also shown as a dotted line. These estimates use the 
$U-V$ CM relation observed in Coma and Virgo as a constraint
(Bower, Lucey \& Ellis, 1992). However, in the top
panel, the F300W$-$F702W color magnitude relation is much steeper
than the prediction for a monolithic stellar population. A least squares
fit to the data points gives a slope of $-0.83$, significantly steeper 
than for a single stellar population ($-0.16$ at $z_F=10$). However,
we should caution our readers by adding in figure~1 a conservative
$4\sigma$ detection limit from the F300W images (shaded line).
The limiting magnitude in F702W ($\simlt 28$) is safely away from the galaxies 
presented in this sample. This implies many faint galaxies have not been 
detected and thus would appear red. Hence, the robust assumption that
we can make with the present data is that there is an increased 

\centerline{\null}
\vskip3truein
\includegraphics{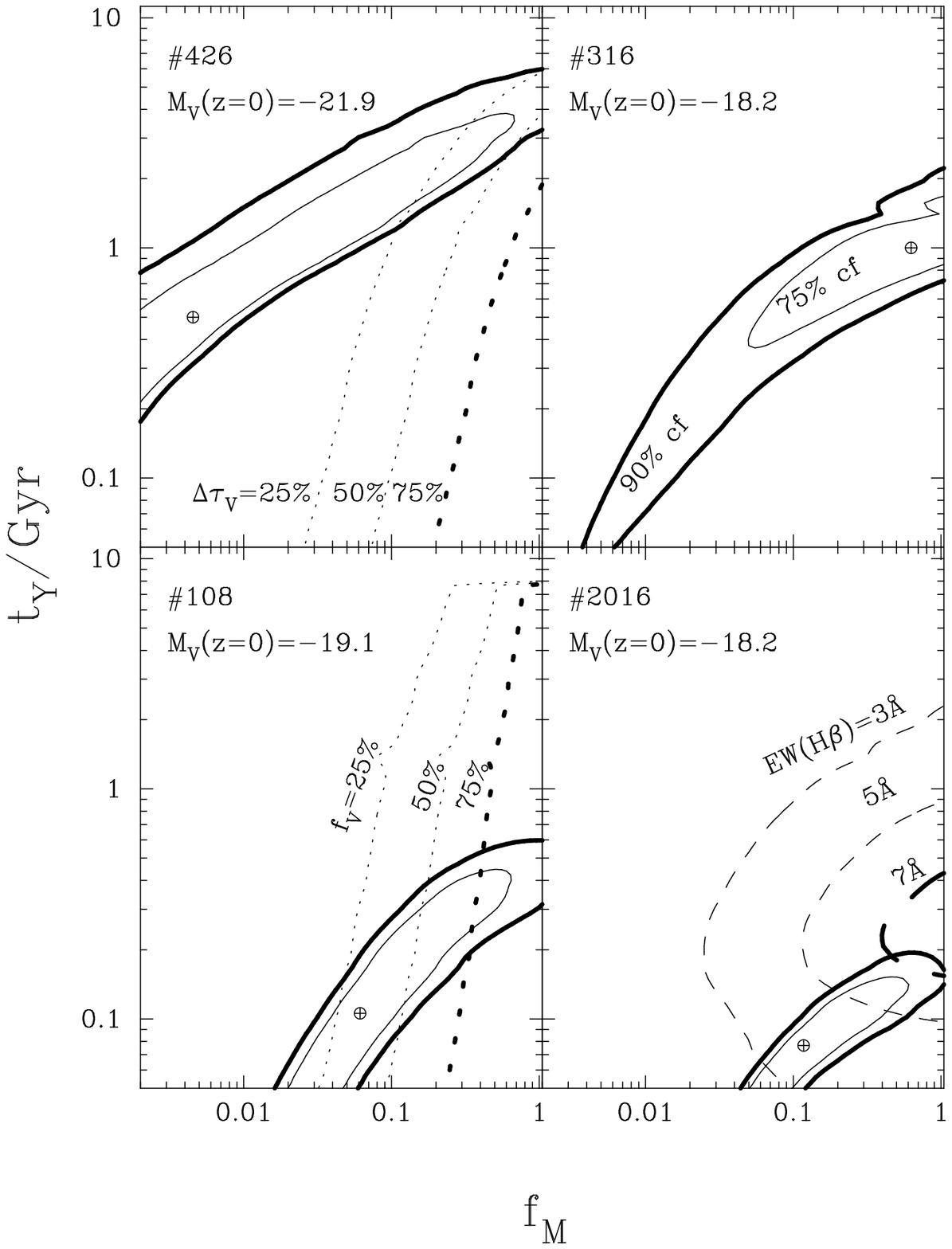}
\figcaption[nuvapj_f2.eps]{Contours at the 75 and 90\% (thick line) 
confidence levels showing the most probable age ($t_Y$) and 
mass fraction ($f_M$) in young stars for four galaxies from the sample.
The dotted lines represent reductions of luminosity-weighted ages 
($\Delta\tau_V$; top left panel) or fractional luminosity contribution 
from young stars in the $V$ band ($f_{V}$; bottom left panel) 
by 25, 50 and 75\% (thick line), respectively.
The bottom right panel shows a rather blue 
early-type system (\#2016), which could be caused either by a small 
($\sim 10\%$) fraction of very young stars up to half of the
stellar mass content of the galaxy in $t\simlt 0.2$ Gyr stars. 
The dashed lines are contours of
fixed $H\beta$ index for solar metallicity, at 3, 5 and 7\AA\
(thick line). 
The galaxies are labelled by their $V$ band absolute 
luminosity at zero redshift, evolved according to the best estimate 
for $t_Y$ and $f_M$, marked by circled crosses.\label{f2}}
\vskip+0.2truein

\noindent
scatter in the faint end of the color-magnitude relation, showing a 
significant fraction of blue early-type systems in NUV-optical 
colors. This excess blueness of the fainter early-type systems 
can only be explained by a younger stellar component. This is a 
feature which goes unnoticed in optical wavebands since the flux of 
small fractions in young stars is negligible compared to the contribution
from the bulk of --- older --- stars in the galaxy. However, the F300W
filter maps rest frame 2000\AA\  which is a spectral range where the flux from 
a large mass in old stars can be overwhelmed by a small fraction of
young stars. 

\section{Near-ultraviolet flux as age estimator}
In order to estimate the fraction of young stars we consider a simple
two-stage star formation history (SFH), where stars are instantaneously formed
at two different epochs (i.e. we add two simple stellar populations).
The old component is formed at a redshift $z_F=3$ which gives an age of
7 Gyr at the redshift of the cluster for a $\Lambda$-dominated flat
cosmology ($\Omega_m=0.3$, $H_0=70$km s$^{-1}$ Mpc$^{-1}$
used hereafter). The age ($t_Y$) and the mass fraction ($f_M$) of the 
young component are explored in a wide range 
($0.05 < t_Y/Gyr < 7.0$; $10^{-4}<f_M<1$). 
The most likely SFH to explain the observations is determined by
a $\chi^2$~test searching the parameter space spanned by
($t_Y$,$f_M$) using several metallicities: 
$Z/Z_\odot = \{ 1/50, 1/20, 0.25, 0.5, 0.75, 1, 1.25, 1.5\}$,
and a range of values for $E_{B-V}$
\footnote{The galactic extinction at the position of the cluster is
$E_{B-V}=0.015$ (Schlegel, Finkbeiner \& Davis 1998).}
between $0.0$ and $0.5$ mag in steps of $\Delta E_{B-V}=0.05$~mag, using 
the standard prescription for starbursts(Calzetti 1997). A more generic 
dust extinction curve such as the $\lambda^{-0.7}$ law from 
Charlot \& Fall (2000) give similar answers, which is to be expected 
if we take into account the spectral range covered by the available 
photometry. The predicted $F300W-F702W$ color for simple stellar 
populations reddened using these two prescriptions differed 
less than 0.1 mag for the age, metallicity and dust range explored 
in this paper. For each choice 
of parameters ($t_Y$,$f_M$,$Z$,$E_{B-V}$) we compute the three colors 
available from the data: ($c_1=$F300W$-$F702W; $c_2=g-r$; $c_3=r-i$)
and perform a $\chi^2$ test comparing the data with the predictions
for a given SFH, metallicity
and dust model using the population synthesis models of 
Bruzual \& Charlot (2000) for a Scalo IMF with mass cutoffs at $0.1$ and 
$100M_\odot$. 
\begin{equation}
\chi^2_{Z,EBV} = \Sigma_i \frac{(c_i-C_{i,Z,EBV})^2}{\sigma_i^2} + 
	\frac{(c^{z=0}_{UV}-C^{z=0}_{UV})^2}{\sigma_{UV}^2},
\end{equation}
where $C_{i,Z,EBV}$ are the predicted colors.
The photometric uncertainties $\sigma_i$ are taken from
the observations for F300W$-$F702W and from Dressler \& Gunn (1992)
for $g-r$ and $r-i$, who estimate their photometric uncertainties
to be less than 10\% down to $r\sim 23$. Taking into account the added
uncertainties from population synthesis models, we assume a 
conservative estimate of $\sigma_i=0.1$ for the optical colors.
The second term in equation (1) is a constraint from local clusters.
For a given choice of parameters, we evolve the galaxy to zero 
redshift and check whether its $U-V$ vs $M_V$ CM relation satisfies
the linear constraint found locally (e.g. Coma and Virgo, 
Bower et al. 1992). This means our estimates are only valid on
the assumption that Cl0939+4713 will evolve into a cluster similar
to the ones found locally. However, relaxing this constraint does not
alter significantly the fraction in young stars, which means
the photometry at $z=0.41$ already hints at an evolution in Cl0939+4713
very similar to local clusters.

\begin{deluxetable}{cccccccccc}
\footnotesize
\tablecaption{Young Stellar Fractions\label{tbl-1}}
\tablewidth{0pt}
\tablehead{ID & Type & F702W & F300W$-$F702W & $M_V(z=0)$ & $t_Y$/Gyr 
& $f_M$ & $\Delta\tau_V$ & $E_{B-V}$ & $Z/Z_\odot$}
\startdata
2016 &  E    & 22.0 & 0.14 & $-$18.2 & 0.08 & 0.118 & 0.57 & 0.00 & 0.50 \cr
 316 &  S0   & 21.6 & 2.67 & $-$18.2 & 1.00 & 0.628 & 0.74 & 0.05 & 0.50 \cr
 291 &  S0   & 21.3 & 1.88 & $-$19.5 & 0.05 & 0.008 & 0.09 & 0.05 & 0.50 \cr
 184 &  E/S0 & 22.0 & 1.47 & $-$18.8 & 0.12 & 0.027 & 0.19 & 0.00 & 0.50 \cr
 172 &  E    & 20.5 & 3.39 & $-$20.2 & 0.85 & 0.107 & 0.28 & 0.20 & 0.25 \cr
 149 &  S0   & 21.1 & 3.23 & $-$19.6 & 1.00 & 0.248 & 0.46 & 0.20 & 0.25 \cr
 303 &  E/S0 & 21.6 & 2.83 & $-$19.3 & 0.36 & 0.035 & 0.15 & 0.15 & 0.25 \cr
 370 &  S0   & 22.2 & 2.52 & $-$18.8 & 0.21 & 0.013 & 0.08 & 0.00 & 0.50 \cr
 108 &  E/S0 & 21.5 & 1.37 & $-$19.1 & 0.11 & 0.061 & 0.34 & 0.15 & 0.25 \cr
 833 &  S0   & 20.5 & 3.38 & $-$20.4 & 0.38 & 0.013 & 0.07 & 0.00 & 0.75 \cr
\enddata
\end{deluxetable}

Table~1 gives the photometry and best estimates found for some of the
early-type systems in Cl0939+4713 which have observed counterparts 
in Dressler \& Gunn (1992) and give a good fit ($\chi^2 < 1.0$). 
The mass fraction in young stars varies widely among galaxies. In order 
to show the range in $t_Y$ and $f_M$ allowed, we show in figure~2 the 
contours at 75 and 90\% (thick line) confidence levels for some of these
galaxies. The crosses give the position of the minima. The elongated
contours are a consequence of the age-mass degeneracy
mentioned above. However, one can see from the figure that for 
some of these galaxies, a significant fraction in young stars is
needed in order to explain their colors. In order to gauge the
contribution from young stars to the photometry of the system, we
define the fractional contribution of the young stellar component to 
the luminosity of the galaxy. Using the $V$ band as reference:
\begin{equation}
f_V = \frac{L_{Y,V}}{L_{O,V} + L_{Y,V}} = \frac{f_M}{(1-f_M)\xi_V+f_M},
\end{equation}
where $\xi_V = \Upsilon_{Y,V}/\Upsilon_{O,V}$ is the ratio of the
mass-to-light ratios between the young and the old components in the
reference band. 

From an observational point of view it is worth relating $f_V$ to 
the fractional age change ($\Delta\tau_V$) between the oldest burst
($t_O$; $z_F=3$) and the $V$ band luminosity-weighted age ($t_V$)
after adding the younger burst ($t_Y$) whose mass fraction 
is given by $f_M$:


\begin{equation}
\Delta\tau_V = \frac{t_O-t_V}{t_O} = \Big( 1-\frac{t_Y}{t_O}\Big) f_V
\end{equation}
The top left panel of figure~2 shows three contours at
$\Delta\tau_V = \{ 25, 50, 75\%\}$, and the bottom left panel shows
similar contours for $f_V$. 

\centerline{\null}
\vskip3truein
\includegraphics{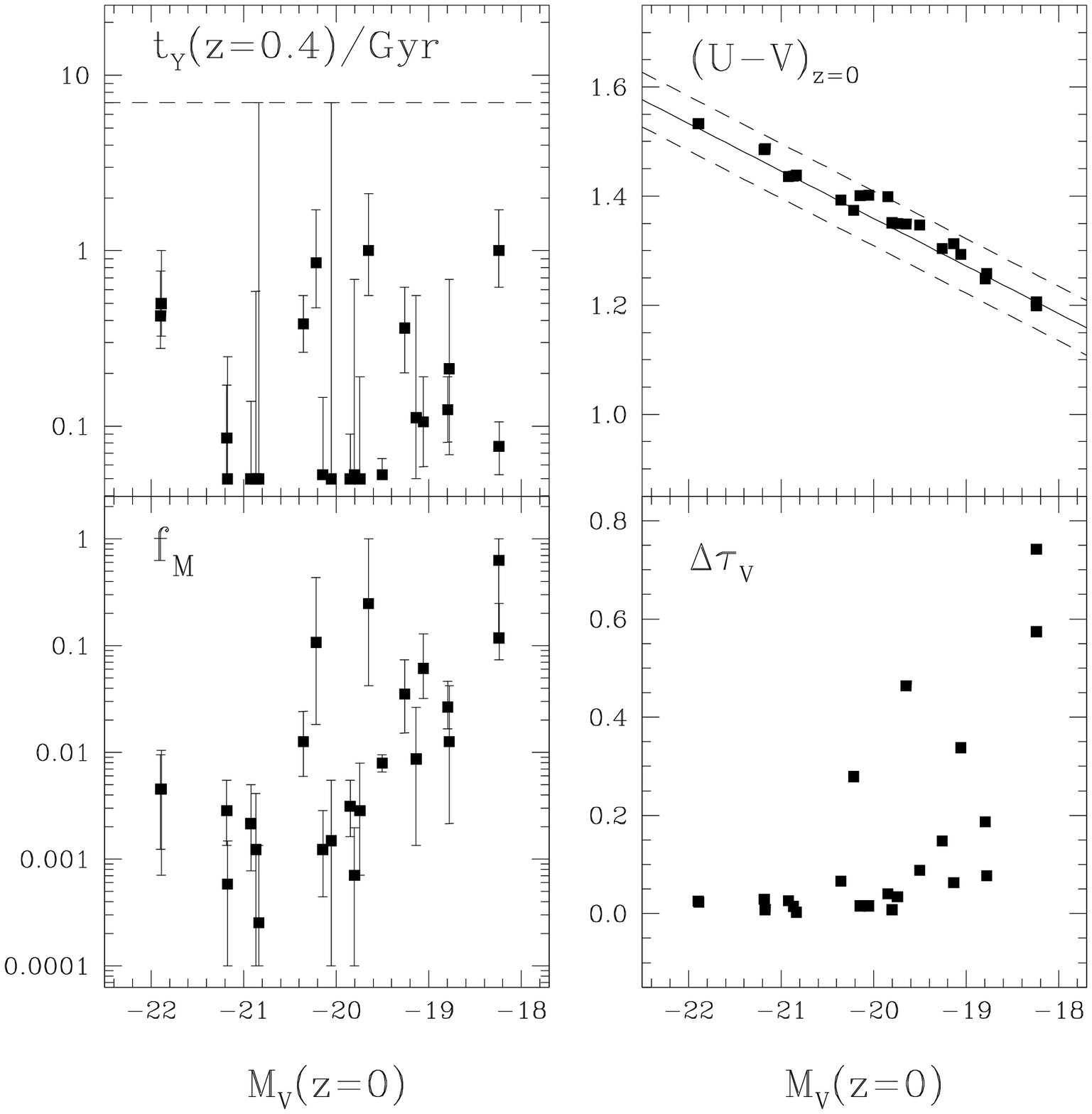}
\figcaption[nuvapj_f3.eps]{Age ({\sl top left}) and mass fraction
({\sl bottom left}) in young stars for the sample of
galaxies analyzed. The results are plotted against 
$V$ band absolute luminosity expected at $z=0$ for the SFH
determined by $t_Y$ and $f_M$. Error bars are shown at the 90\%
confidence level at fixed $f_M$ ({\sl top left}) or fixed $t_Y$
({\sl bottom left}). The dashed line in the top left 
panel shows the age of a monolithic formation scenario at 
$z_F=3$. The top right panel is the predicted $U-V$ vs $M_V$ 
color magnitude relation at $z=0$. The solid and dashed lines 
are the best linear fit and scatter of the observed CM relation 
in Coma (Bower et al. 1992). The bottom right panel gives the 
fractional age reduction ($\Delta\tau_V$) between a monolithic 
scenario at $z_F=3$ and the inferred luminosity-weigthed age.
\label{f3}}
\vskip+0.2truein

\noindent
For some of the
galaxies, the most probable SFH corresponds to a fractional age
reduction (with respect to the old population) above 50\% with 
respect to a single burst at $z_F=3$, even 
though these galaxies will display at $z=0$ a similar CM 
relation as the one found in Coma. The bottom right panel overlays three contours
corresponding to equivalent widths of $H\beta$  of 3, 5, 
and 7\AA\ (thick), showing that a combined NUV-optical photometry 
along with a spectroscopic measurement of Balmer indices will greatly 
reduce the age-mass degeneracy.


\section{Discussion}
This paper represents a first attempt at {\sl quantifying} the fraction
in young stars in early-type cluster galaxies. From a spectrophotometric
point of view, these galaxies commonly feature a population of very 
old stars, without any traces of recent star formation. 
However, from a dynamical point of view,
early-type systems must form from succesive merging stages. The 
observation of these mergers locally display an active process
of star formation, but progenitors with very little cold gas will
undergo mergers which will not trigger star formation
(Van Dokkum et al. 1999). So far, observations of elliptical galaxies
in the NUV spectral range were confined to the search of very
old helium core-burning stars as a way of estimating very old ages in
stellar populations. However, the existence 
of early-type systems with a clear E+A spectral feature show that 
galaxies with this morphological type may have a more interesting 
star formation history. We focus on
the NUV spectral range around rest frame 2000\AA\  where young
stars overwhelm any contribution from older stellar populations.
Observations of spectral indices have found a rather large age 
spread in field and group early-type systems (Trager et al. 2000).
However, their analysis --- based on simple stellar populations
--- cannot quantify a young-to-old mass fraction. A comparison 
between NUV and optical colors allows us to quantify
this stellar mass fraction, although with large error bars.
Figure~3 gives the age and mass fraction of young stars ({\sl left}) in 
early-type galaxies in cluster Cl0939+4713. Notwithstanding the
large error bars, there is a significant trend towards younger 
ages and mass fractions in fainter galaxies, with a large fractional
age difference with respect to a monolithic star formation history
at high redshift which can be quantified by a luminosity-weighted age
fractional change ({\sl bottom right}). However, these galaxies can 
still accomodate the tight CM relation found in local clusters 
({\sl top right}). In the framework of current models of star
formation, one could accomodate this trend with lower 
star formation efficiencies in fainter galaxies, also resulting 
in a significant change with redshift of the slope of the correlation
between  mass-to-light ratio and mass (Ferreras \& Silk 2000). This would 
imply that in the latest merging stages the less massive
early-type systems could still undergo some star formation, whereas
more massive galaxies would just merge stars and hot gas, without
triggering star formation. The fact that the method presented here
is very  sensitive to small fractions of young stars poses a strong 
constraint on recent star formation in massive ellipticals. We
should emphasize that our claim only applies to the scatter 
at faint magnitudes, since red faint galaxies might lie beyond
the detection limit of the archival F300W images used here. 
There are two possible alternatives to explain the blueness
of these faint early-type systems. However, one can argue
against them for the following reasons:
\begin{itemize}
\item[$\bullet$] Old stellar populations could contribute 
to wavelengths around rest frame 1500\AA\ , however this is negligible
in comparison to the flux from $t_Y\simlt 1$ Gyr stars. A simple stellar 
population at the redshift of the cluster ($z=0.41$) does not yield
colors F300W$-$F702W$<4$ for ages above $7$ Gyr ($z_F>3$) 
both for a Scalo and a Salpeter IMF. Exploring these two IMFs allows us to
estimate the contribution of old low-mass stars to the NUV,
the contribution in low mass stars being significantly larger for
a Salpeter IMF. Yet, the F300W$-$F702W colors did not differ by more
than $0.1-0.2$ mag. Furthermore, we compared the output of the 
population synthesis model used here (Bruzual \& Charlot 2000) with 
the models of Yi et al. (1999) who take special care in computing 
the NUV contribution from core helium-burning stars. The only 
significant discrepancy is found for ages above 15~Gyr, which 
will be important in the study of local elliptical galaxies.
However, the redshift of the cluster studied here ($z=0.41$)
sets an upper limit to the age of any stellar population between
9 and 10 Gyr for reasonable cosmologies. Hence, even though
the predicted fractional mass distribution in young stars
may slighty differ among models, the connection between 
an increased scatter in the faint end of the NUV$-$optical 
color-magnitude relation and recent star formation is fairly
robust.
\item[$\bullet$] Another possibility for such blue colors would be 
extremely low metallicities. However these metallicities predict optical
$g-r$ and $r-i$ colors which are too blue with respect to the
observations. In fact a simple stellar population 
with very low metallicity ($Z_\odot/50$)
formed at $z_F=5$ gives F300W$-$F702W$=2.7$, whereas the optical
colors are blue as well: $r-i=0.3$. Furthermore, the evolution to $z=0$
gives a color $U-V=0.6$ that is hard to reconcile with local clusters.
\end{itemize}
We believe the analysis of $\lambda\sim 2000$\AA\ NUV light 
in moderate and high redshift galaxies
should find its place in the combined effort 
to search for the connection between the dynamical and the 
spectrophotometric histories of galaxies, along with other currently 
used age estimators such as Balmer indices or mass-to-light ratios.

\acknowledgments
We are indebted both to St\'ephane Charlot and Sukyoung  Yi for useful 
suggestions and for making available their latest population 
synthesis models. IF is supported by a 
grant from the European Community under contract HPMF-CT-1999-00109


\end{document}